
\input phyzzx
\def \fff {\vrule width 0.5pt height 5pt depth 1pt}
\def \pp { {=\hskip -3.75pt {\fff}\hskip 3.75pt}}
\def \np {Nucl. Phys.}
\def \pl {Phys. Lett.}
\def \cmp {Comm. Math. Phys.}
\def \cqg {Class. Quantum Grav.}
\REF \z {B. Zumino, \pl \ \underbar{87B} (1979), 205.}
\REF \agf {L. Alvarez-Gaume and D. Z. Freedman, \cmp\  \underbar{80}
(1981), 443.  T. Curtright and D. Z. Freedman, \pl\
 \underbar{90B} (1980), 71.}
\REF \ghr {S. J. Gates, C. M. Hull and M. Rocek, \np\
 \underbar{B248} (1984),
157.}
\REF \hp {P. S. Howe and G. Papadopoulos, \np\ \underbar{B289} (1987), 264;
\cqg\ \underbar{5} (1988), 1647.}
\REF \vkpr {P. di Vecchia, V. Knizhnik, J. Petersen and P. Rossi, \np\
\underbar{B253} (1985), 701.}
\REF \ht {P. S. Howe, S. Penati, M. Pernici and P. K. Townsend, \cqg\
\underbar{6} (1989), 1125}
\REF \gp {G. Papadopoulos, \cqg\ \underbar{7} (1990), 239.}
\REF \paho {P.S. Howe and G. Papadopoulos, Phys. Lett.\ \underbar{263B}
    (1991), 230; Phys. Lett.\ \underbar{267B} (1991), 362}
\REF \cmh {C. M. Hull, \pl\ \underbar{178B} (1986), 357;
in Super Field Theories, Proc. Nato Workshop,
 Vancouver (1986), ed. H. C. Lee,
 V. Elias, G. Kunstatter, R. B.
 Mann and K. S. Viswanathan (Plenum, New York 1987.}
\REF \wn {B. de Wit and P. van Nieuwenhuizen,
\np\ \underbar{B312} (1989), 58.}
\REF \dn {G. W. Delius and P. van Nieuwenhuizen in String's '89,
 Proc. of the
superstrings workshop Texas A\&M university (1989),
 ed. R. Arnowitt, R. Bryan,
M. J. Duff, D. Nanopoulos and C. N. Pope (World Scientific, London 1989).}
\REF \az {A. B. Zamolodchikov, Teor. Mat. Fiz.\ \underbar{65} (1985), 1205.}
\REF \cmhB {C. M. Hull, \pl\ \underbar{240B} (1990), 110; "Higher Spin
Conformal Symmetries and W-Gravities" QMW/PH/9 (1990).}
\REF \gr {G. de Rham, Comm. Math. Helv.\ \underbar{26} (1952), 328.}
\REF \mb {M. Berger, Bull. Soc. Math. France\ \underbar{83} (1953), 279.}
\REF \fn {A. Frolicher and
 A. Nijenhuis, Proc. Ned. Akad. van Wet. Amsterdam
(series A)\ \underbar{59} (1956), 338; \underbar{59} (1956), 540.}
\REF \will {T. J. Willmore, Journ. London Math. Soc.\
 \underbar{35} (1960), 425.}
\REF \s {W. Slebodzinski, Colloq. Math.\ \underbar{13} (1964), 49.}
\REF \wB {T. J. Willmore, Journ. London Math. Soc.\
 \underbar{43} (1968), 321.}
\REF \bw {J. Bagger and E. Witten, \np\ \underbar{B222} (1983), 1.}
\REF \rb {R. L. Bryant, Ann. of Math.\ \underbar{129} (1987), 525.}
\REF \gris {M. T. Grisaru, A. E. M. Van de Ven and D. Zanon, \pl\
\underbar{173B} (1986), 423; \np\ \underbar{B277} (1986), 388 and 409.}

\date={December 1991}
\pubtype={ }
\titlepage
\title { Holonomy Groups and W-symmetries}
\author {P.S. Howe,}
\address { Department of Mathematics,
\break King's College London, \break Strand, \break London WC2R 2LS.}
\andauthor {G. Papadopoulos,}
\address {Physics Department, \break
Queen Mary and Westfield College,
 \break Mile End Road, \break London E1 4NS.}

\abstract {Irreducible sigma models, i.e. those for which the partition
function does not factorise, are defined on
 Riemannian spaces with irreducible
holonomy groups.  These special geometries are characterised by the existence
of covariantly constant forms which in turn give rise to symmetries
of the supersymmetric sigma model actions.  The Poisson bracket
algebra of the corresponding currents is a W-algebra. Extended
supersymmetries arise as
special cases.}

\chapter {Introduction}

It has been known for many years that the geometry of the target space of two
dimensional sigma models is restricted when there are further supersymmetries;
in particular, N=2 supersymmetry requires that the target space be a K\"ahler
manifold [\z], and N=4 supersymmetry requires that it be a hyperk\"ahler
manifold [\agf].  More exotic geometries arise in heterotic sigma models
with torsion and in one-dimensional models [\ghr, \hp, \cmh].  More recently it
has been realised that sigma models can admit further symmetries which are
non-linear in the derivatives of sigma model field.  The prototype of this
type of symmetry is the non-linear realisation of supersymmetry using free
fermions [\vkpr]; further instances have been given in the context of
supersymmetric particle mechanics [\ht, \gp] and in N=2 two-dimensional models,
where it has been realised that it is not necessary to impose the vanishing of
the Nijenhuis tensor [\cmh, \wn]. In [\paho] a preliminary investigation into
non-linear symmetries of other two-dimensional supersymmetric sigma models
was presented.
A related type of symmetry occurs in bosonic
sigma models, the so-called W-symmetry [\az, \cmhB].

In this article we combine the issues of the geometry of the target spaces and
the non-linear symmetries of two dimensional supersymmetric sigma models.  In
the case of N=2 and N=4 supersymmetries, for example, the additional
structures on the (Riemannian) target spaces reduce the holonomy groups from
$O(n)$ to $U({n \over 2})$ and $Sp({n \over 4})$ respectively where n=dim M, M
being the target space.  We shall investigate the symmetries associated with
other holonomy groups, restricting our study to manifolds which are not
locally symmetric spaces and which have irreducible holonomy groups.
Irreducibility here means that the n-dimensional representation of $O(n)$
remains an irreducible representation of the holonomy group $G\subset O(n)$. In
the  case that the connection is the Levi-Civita connection, the
irreducibility of the holonomy implies that M is an irreducible
Riemannian manifold (if $\pi_1(M)=0$)\ [\gr], i.e. M is not a
product $M_1\times M_2\times \cdots $
such that
the metric can be written as a
direct sum with each component
depending only on the co-ordinates of
the corresponding factor of the target manifold.  In
field-theoretic terms, sigma models on metrically
reducible spaces factorise
into sigma models on factor spaces in the sense
that the partition function
factorises.  However, interesting
symmetries can arise on reducible manifolds
in which the factors transform into each other; an example
of this behaviour
occurs in the case of W-symmetry where the
target spaces are reducible (for non-locally symmetric spaces).

The irreducible holonomy groups associated with Levi-Civita connections on
Riemannian manifolds have been classified by
Berger [\mb].  The possible holonomy groups that can arise are $SO(n)$,
$U({n \over 2})$, $SU({n \over 2})$, $Sp({n \over 4})$ and $Sp(1) \cdot
Sp({n \over 4})=Sp(1)\times_{Z_2}  Sp({n \over 4})$ together
with the exceptional cases,
$G_2$ (n=7) and $Spin(7)$ (n=8).  In each case there is an associated
covariantly constant (with respect to the Levi-Civita connection) totally
antisymmetric tensor, and it is this fact which implies the existence of an
associated symmetry of the corresponding supersymmetric sigma model
[\dn, \paho].  We call such Riemannian geometries special.
This classification is not strictly
applicable  to models with torsion for which the corresponding analysis has
not been done. Nevertheless , irreducible holonomy
is a useful retriction to impose and the
covariantly  constant tensors are the same as in the torsion-free case. In
many cases of  interest we shall in any case set the torsion to zero. Indeed,
for both the exceptional cases, $G_2$ and $Spin(7)$, it turns out that the
torsion must vanish. The Riemannian (i.e. torsion-free) case is the most
interesting one from the point of view of the algebraic structure of the
non-linear symmetries under consideration, since in this case, as we shall
show, the corresponding currents, together with the (super) energy-momentum
tensor, generate super W-algebras via Poisson Brackets. These algebras are
extensions of the (classical) superconformal algebra by additional currents
which are, in general, of higher spin. It is of interest to note that the
field theory models which provide realisations of classical W-algebras
presented here are highly non-trivial field theories. This fact makes the
analysis of the corresponding quantum algebras more complicated and we shall
not pursue this topic in this paper.

In section 2 we discuss the general form of symmetries generated by
covariantly constant antisymmetric tensors.
At the classical level these symmetries are of
semi-local (superconformal) type, i.e. the parameters depend on some,
but not all, of the coordinates of superspace,
and in general generate an infinite number of symmetries of this type.
However, there are examples of finite dimensional semi-local symmetry
algebras, for example on
manifolds with $G=SO(n)$. In some cases it is possible to get
a finite-dimensional Lie algebra by restricting the parameters to be constant;
an example
of this type is given by Calabi-Yau manifolds ($G= SU({n \over 2})$).
When the torsion vanishes, as we remarked above, we obtain finite-dimensional
W-algebras, i.e. W-algebras generated by a finite number of currents.
For some purposes it is useful to
regard the invariant antisymmetric tensors  associated with the special
geometries as vector-valued forms, and we include in this section a
brief review of the way
such vector-valued forms give rise to derivations of the algebra of
differential forms on the target space [\fn, \will].  In section 3, we
introduce
Poisson Brackets and compute them for the currents of the type we are
interested in. In section 4 we study the
various cases listed above, and in section 5 we make some concluding remarks.

\chapter {General Formalism}

Let $\Sigma$ denote the (1,0) (or N=1)
superspace extensions of two-dimensional Minkowski space, with real
light-cone co-ordinates $(y^{\pp}, y^=, \theta^+)$ (resp $ (y^{\pp},
 y^=,  \theta^+,$  $\theta^-))$.  The supercovariant derivatives
$D_+\  (D_+, D_-)$ obey
$$ D_+^2=\ i\ \partial_{\pp} \eqn\aone$$
and
$$D_+^2=\ i\ \partial_{\pp};\  D_-^2=\ i\ \partial_=;\ \
   \  \{D_+,D_-\}=0  \eqn\atwo$$
respectively.  Let ($M,g$) be a Riemannian target space (metric $g$) equipped
if necessary with a closed three-form $H=3db$, where $b$ is a locally defined
two form.  Local co-ordinates on $M$ will be denoted $x^i, i=1,...n$, and
the sigma model superfield by $X^i$.  The (1,0) action is
$$S=\int d^2y d\theta^+\  (g_{ij}+b_{ij})\  D_+X^i\  \partial_=X^j
\eqn\athree$$
and the (1,1) action is
$$S=\int d^2y d^2\theta\  (g_{ij}+b_{ij})\  D_+X^i\  D_-X^j \eqn\afour$$

Let $\omega_L$ be an $(l+1)$-form on M
$$\omega _L=L_{i_1\cdots i_{l+1}}\  dx^{i_1}\wedge \cdots \wedge dx^{i_{l+1}}.
\eqn\afive$$ We introduce a vector-valued $l$-form, $L^i$, and a $Lie(O(n))$
valued  $(l-1)$-form ${\cal L}^i_j$ by defining
$$L^i=L^i_L\  dx^L \eqn\asix$$
$${\cal L}^i_j= L^i_{jL_2}\  dx^{L_2}
				\eqn\aseven$$
where
$$dx^L:=dx^{i_1} \land \cdots \land dx^{i_l}
			\eqn\aeight$$
$$dx^{L_2}:=dx^{i_2} \land \cdots \land dx^{i_l}
		\eqn\anine$$

If $\omega_L$ is covariantly constant, i.e. if
$$\nabla_j^{(+)} L_{i_1...i_{l+1}}=0  \eqn\aten$$
with
$${\Gamma^{(\pm)}}^ i_{jk}=\Gamma^i_{jk} \pm {1\over 2}{H^i}_{jk}
  \eqn\aeleven$$
then the transformation
$$\delta_L X^i =\ a_{-l}\  L^i_L\  D_+X^L
		\eqn\atwelve$$
where
$$D_+X^L:=D_+X^{i_1}\cdots D_+X^{i_l}
						\eqn\athirteen$$
is a symmetry of the (1,0) action if the parameter $a_{-l}$ satisfies
$\partial_= a_{-l}=0$ and a symmetry of the (1,1) action if $D_-a_{-l}=0$.
The notation for the parameter indicates that it has Lorentz weight ${-l\over
2}$ and is thus Grassmann even or odd
 according to whether $l$ is an even or odd
integer. The (1,1) action is also invariant under
$$\delta_L X^i =\ a_{+l}\  L^i_L\  D_-X^L  \eqn\afourteen$$
if
$$\nabla_j^{(-)} L_{i_1...i_{l+1}}=0  \eqn\afifteen$$
and $D_+a_{+l}=0$.

The above symmetry transformations are associated with derivations of the
algebra of forms, $\Omega$ on $M$.  Let $\Omega_p$ denote the space of
p-forms, so that $\Omega=\oplus^n_{p=0} \Omega_p$, and $\Omega^1_l$ the
space of vector-valued $l$-forms.
 We recall that a derivation $D$ of degree r
satisfies the following properties:
$$\eqalign {&
a)\qquad    D(a\omega+b\rho) =aD\omega+bD\rho;\qquad a,b\in \bf {R}
\qquad \ \ \ \ \ \ {\rm Linearity}
\cr &
c)\qquad    D\Omega_p\subset \Omega_{p+r}; \qquad \ \ \ \ \ \ \ \ \ \ \ \ \ \
\ \ \ \ \ \
{\rm Degree}\  r
\cr &
d)\qquad   D(\omega \wedge \rho) = D\omega \wedge \rho +(-1)^{pr}\omega \wedge
D\rho,\ \   \omega\in \Omega_p; \qquad  {\rm Leibniz\ property} \cr}
\eqn\asixteen $$

The commutator of two derivations $D_r$ and $D_s$ of degrees r and s
is defined by
$$[D_r,D_s]:= D_r D_s - (-1)^{rs} D_s D_r  \eqn\aseventeen$$
and the Jacobi indentity
$$[D_r,[D_s,D_t]] + (-1)^{t(r+s)} [D_t,[D_r,D_s]] + (-1)^{r(s+t)}
[D_s,[D_t,D_r]]=0 \eqn\aeighteen$$
holds for any three derivations.  Thus the space of derivations is a $\bf
Z$-graded super Lie algebra.

There are two types of derivation both of which are defined by
vector-valued forms.  If $v$ is a vector field, i.e. a vector-valued 0-form,
then the interior product of $v$ with a $p$-form, denoted $\iota_v\omega$
is a derivation given by
$$\iota_v \omega = p\ v^i\omega_{iP_2} dx^{P_2}\eqn\crazya
$$
Since $d$ is also a derivation
 we can generate another one from its commutator
with $\iota_v$,
$$
\iota_v d + d\iota_v = d_v\eqn\crazyb
$$
This is just the Lie derivative, normally denoted as ${\cal L}_v$. A  similar
construction can be carried out for a general vector-valued form.
If $L\in \Omega^1_l$ and $\omega\in \Omega_p$ we
define their interior product $\iota_L\omega$ by
$$
\iota_L\omega:=\  p\  \omega_{iP_2}\  L^i_L\  dx^L\wedge dx^{P_2}
\eqn\anineteen$$
Another notation for this construct
 is $\omega\bar {\wedge} L$ [\will]; we shall
use both. It is easy to check that $\iota_L$ is a derivation. Taking the
commutator of $\iota_L$ with $d$ we get a new derivation $d_L$ which
generalises the Lie derivative,
$$
\iota_L d + (-1)^l d\ \iota_L = d_L \eqn\crazyc
$$
It has the property that it commutes with $d$, $d_L d =(-1)^l d d_L$,
and is determined by its action on $\Omega_0$,
$$d_L f=df \bar {\wedge} L  \eqn\atwtwo$$
On a p-form $\omega$,
$$d_L \omega =d \omega \bar {\wedge} L+(-1)^l d(\omega \bar {\wedge} L)
\eqn\atwthree$$
The Nijenhuis tensor (concomitant) $[L,M]$ of two vector-valued
forms $L$ and $M$ of degrees $l$ and $m$ is defined by
$$[d_L,d_M]= d_{[L,M]}.  \eqn\atwfour$$
The Nijenhuis tensor can be worked out by observing that
$$d_L x^i =L^i  \eqn\atwfive$$
so that
$$\eqalign {
[d_L,&d_M] x^i
=  [L,M]^i \cr
&=
dL^i\bar {\wedge} M  +  (-1)^l\  d(L^i\bar {\wedge} M)  -
 (-1)^{lm}\  \big{(}dM^i \bar {\wedge} L  +  (-1)^m d(M^i \bar {\wedge} L)
\big{)}  \cr }                                        \eqn\atwsix$$
where $L^i$ is regarded as an $l$-form for each value of $i$.
In more detail,
$$\eqalign{
[L,M]^i &= [L,M]^i_{LM} dx^L \wedge dx^M
\cr &
=
\big {(}L^j_L\  \partial_jM^i_M - M^j_M\ \partial_jL_L - l\  L^i_{jL_2}\
\partial_{l_1}\ M^j_M  + m\  M^i_{jM_2}\  \partial_{m_1}L^j_L\big{)}\  dx^{LM}.
\cr }
 \eqn\atwseven$$
It is straightforward to verify that $[I,I]$ is the usual Nijenhuis
tensor, $N(I)$, for the case $L=M=I$,
 an almost complex structure.  Hence the
integrability condition for an almost complex structure to be complex,
 $N(I)=0$, is
equivalent to the condition $d^2_I=0$.

The other commutators are
$$[\iota_L,d_M] = d_{M \bar {\wedge} L} + (-1)^m \iota_{[L,M]}  \eqn\atweight$$
and
$$[\iota_L,\iota_M]= \iota_{M \bar {\wedge} L} + (-1)^{l+m+lm}
\iota_{L \bar {\wedge} M}
\eqn\atwnine$$
where
$$(M \bar {\wedge} L)^i:=\  m\  {M^i}_{jM_2}\  L^j_L\  dx^L \wedge dx^{M_2}
\eqn\athirty$$

 From the Jacobi identity one can
 derive a number of identities for the tensors
which arise in the commutators, for example,
$$[L,[M,N]]+(-1)^{n(l+m)}[N,[L,M]]+(-1)^{l(m+n)}[M,[N,L]]=0  \eqn\athone$$
and
$$\eqalign {
[L\bar {\wedge} M,N] + (-1)^{(m+1)l}& [L,N\bar {\wedge}M]
-[L,N]\bar {\wedge}M =
\cr &
=
(-1)^{n (l+1)} L\bar {\wedge}[M,N]+
(-1)^{l+1} N\bar {\wedge}[M,L]\cr}    \eqn\athonea$$

We can now compute the commutator of
two transformations of the type (2.12).  It is
$$ [\delta_L,\delta_M] X^i=\delta_{LM}^{(1)} X^i + \delta_{LM}^{(2)} +
\delta_{LM}^{(3)}      \eqn\athtwo$$
where
$$ \delta^{(1)}_{LM}=\
 a_{-m}\  a_{-l}\  [L,M]^i_{LM}\  D_+X^L\  D_+X^M$$
$$\eqalign {
\delta^{(2)}_{LM}
=& a_{-m}\  D_+a_{-l}\  (M \bar {\wedge} L)^i_{LM_2}\  D_+X^L\  D_+X^{M_2}-
\cr &
-\
 a_{-l}\  D_+a_{-m}\  (L \bar {\wedge} M)^i_{ML_2}\  D_+X^M\  D_+X^{L_2}
\cr }$$
and
$$\eqalign {\delta^{(3)}_{LM}=&\
i\  l\ m\  (-1)^l\  a_{-m}\  a_{-l}\  \big{(}{\cal L}^i_j \wedge {\cal M}^j_k +
\cr &
+
 (-1)^{(l+1) (m+1)}
{\cal M}^i_j \wedge {\cal L}^j_k \big{)}_{L_2M_2}\  \partial_{\pp}X^k\
 D_+X^{L_2}\  D_+X^{M_2}\cr }  \eqn\athtwob $$
In general the three terms on the
 right hand sight of (2.33) are not symmetries
by themselves, so that a much larger and more complicated algebra of
transformations will be generated.

In the case that the torsion vanishes
 it is straightforward to show that [L,M]
also vanishes, given that L and M are covariantly constant.
For (1,1) models it is straightforward to show that the left
and right transformations (2.12) and (2.14) commute up to the equations
of motion.

\chapter {Poisson brackets}

Let $\{j_A\}$ be the currents of
 a set of symmetries of a two-dimensional field
theory.  The Poisson Bracket algebra
$$  \{j_A, j_B\}_{PB}= P_{AB} (\{j_A\})            \eqn\aaone$$
of these currents forms a W algebra provided
 that $P_{AB}$ is a polynomial in
the currents $\{j_A\}$ and their derivatives.

The currents of the symmetries \atwelve\ of the action \athree\ (or \afour )
are
$$  j_L= {1\over l+1}\  L_{i_1 \cdots i_{l+1}}\
 D_+X^{i_1}\cdots D_+X^{i_{l+1}}
						\eqn\aatwo$$
These currents are conserved, $D_- j_L =0$ (or $\partial_= j_L =0$),
subject to the eqns. of motion of the action \athree\ (or \afour ).  The
form $\omega _L$ \afive\ satisfies the eqn. \aten.

To get a complete set of currents, it is necessary to include the (super)
energy-momentum tensor $T$, given by
$$ T=\  g_{ij}\  D_+X^i\  \partial_{\pp}X^j.
					\eqn\aathree$$
$T$ generates left-handed supersymmetry transformations and translations.

In the rest of this section we shall assume that the torsion vanishes,
$H=0$, and we shall also focus only on left-handed currents having the
form (3.2) or (3.3); any dependence on the right-handed co-ordinates $
(y^=,\theta^-)$ will be suppressed.

To calculate the Poisson brackets of the currents $j_L$ \aatwo, we
introduce the Poisson bracket
$$\{D_+X^i(z_1), D_+X^j(z_2)\} =\  g^{ij}\ \nabla_{+1} \delta (z_1, z_2)
					\eqn\aafour$$
where $z=(y^{\pp}, \theta^+)$.  This
 Poisson bracket is constructed from
light-cone considerations where the co-ordinate $y^=$ of the flat
superspace is taken as ``time''.

Next we define the ``smeared'' currents $j_L(a_l)$ by
$$ j_L(a_l) = \int dy^{\pp} d\theta^+\  a_{-l}\  j_L   \eqn\aafive$$
where $a_l$ is a function of $z$ with Grassmannian parity $(-1)^l$.
The Poisson bracket of two currents of the form \aafive\ is
$$\eqalign {
\{ j_L(a_{-l}), &j_M(a_{-m}) \}_{PB} =
\cr &
\  (-1)^{l m+m+1}\
		\big{(}{l+m \over l+1}\  j_{L\bar {\wedge} M}(D_+a_{-l}\
a_{-m})\  +\  \bar {j}_{L,M}(a_{-l}\  a_{-m})\big{)}  \cr  }
				\eqn\aasix$$
where
$$ j_{L\bar {\wedge} M} (D_+a_{-l}\  a_{-m}) = {(-1)^{lm} \over l+1} \int
dy^{\pp} d\theta^+\  D_+a_{-l}\  a_{-m}\  (\omega_L\bar {\wedge} M)_{LM}
			\  D_+X^{LM}	\eqn\aaseven$$
and
$$\bar {j}_{L,M}(a_{-l}\ a_{-m}) =
- {(-1)^m\over l} \int dy^{\pp} d\theta^+\  i\
\   a_{-l}\  a_{-m}\  (L \bar {\wedge} M)_{jL_2 M}\
\partial_{\pp}X^j\  D_+X^{L_2 M}.       \eqn\aaeight$$

In the examples we shall see that $\bar {j}_{L,M}$ can be written as a
product of the original currents,
 the energy-momentum tensor $T$ and their
derivatives.

The Poisson bracket of $T$ with $j_L$ is
$$\{T(a_=), j_L(a_{-l})\}_{PB} = (l+1)\ j_L\big {(}\partial_{\pp} a_=\  a_{-l}\
+\ 2\  a_=\  \partial_{\pp}a_{-l}\big {)}\  +\  {1\over i}\ j_L(D_+(D_+a_=\
a_{-l}))\eqn\aanine$$
This formula reflects the fact that $j_L$ has Lorentz weight ${1\over2}(l+1)$.

\chapter {Applications}
\section {$SO(n)$}

The simplest case to analyse is $SO(n)$.  The corresponding invariant
tensor is the $\epsilon$-tensor, $\epsilon_{i_1...i_n}$.
The symmetry transformation is
$$\delta X^i=\  a_{1-n}\  \epsilon^i\ _{j_1...j_{n-1}}\
D_+X^{j_1}\cdots D_+X^{j_{n-1}} \eqn\bone$$

This is a bosonic symmetry for n odd, and it is easy to see that the
commutator of two such transformations is zero.  Comparing with (2.33), we
observe that the first and third terms in the right-hand side vanish
automatically when L=M and the symmetry is bosonic.  The second term is
trivially zero for $n\geq 5$, and for $n=3$ can been seen to be zero by a
short explicit computation.  When $n$ is even (3.1) defines a fermionic
symmetry which is also nilpotent, except in the case $n=2$.  The first and
second terms on the right-hand side of (2.33) vanish trivially unless
$n=2$ or $4$.  In the case $n=4$, the properties of the $\epsilon$-tensor
imply  that both terms are again zero.  In the case $n=2$,
the $\epsilon$-tensor defines an almost complex structure
on $M$ which is integrable (the torsion vanishes
identically); hence, the first term in (2.2) is zero, the second is a
first supersymmetry transformation and the third is a translation.

Thus, for $n\geq 3$, sigma models with $SO(n)$ holonomy can be
characterised by the existence of an Abelian (super)conformal symmetry.
In the case $n=2$, this becomes N=2 (or (2,0) supersymmetry).

\section {$U({n \over 2})=U(m)$}

In this case the antisymmetric tensor is derived from an almost complex
structure ${I^i} _j$, $I^2=-1$.  Models with $U(m)$ holonomy have been
extensively studied in the literature, including the case where $I$ is
not complex, i.e. $N(I) \not= 0$ [\cmh, \wn, \dn].  The commutator of two
transformations defined by $I$ is
$$\eqalign {
[\delta_{I},\delta'_{I}] X^i &
=\
a'_{-1}\  a_{-1}\  N^i_{j_1j_2}\  D_+X^{j_1}\  D_+X^{j_2}\  +\
\cr &
+
\ D_+(a'_{-1} a_{-1})\  D_+X^i\  +\  2i\  (a'_{-1}\  a_{-1})\
\partial_{\pp} X^i}  		\eqn\btwo$$
The second and third terms correspond to first supersymmetry transformation
 and translations, while the Nijenhuis tensor term defines a new symmetry of
the type (2.12).  Since the second and third terms are symmetries by
themselves, so is the first term and this implies that $N_{ijk}$ must be
totally antisymmetric and covariantly constant.

One can now investigate the algebra generated by $\delta_I$, i.e. compute
$[\delta_I,\delta_N]$, etc..  Referring again to equation (2.33), the third
term on the right hand side can be shown to vanish by virtue of the
identity
$${I^i}_l\  {N^l} _{jk}\  +\  {N^i}_{jl}\  {I^l}_k\  =\  0  \eqn\bthree$$
The second term gives a contribution
$$[\delta_I,\delta_N]X^i =
- (a_{-1}\ D_+ a_{-2}\  +\  2\  a_{-2}\  D_+ a_{-1})\
{\hat {N}^i}_{jk}\  D_+X^j D_+X^k +\cdots
				  \eqn\bfour$$
where $\hat {N} = I\bar {\wedge} N$.  $\hat {N}_{ijk}$ is again totally
antisymmetric and covariantly constant so we have a new symmetry of type
(2.12).

Finally the first term gives rise to a transformation involving the
Nijenhuis concomitant of $I$ and $N$, $[I,N]$.  This is the Slebodzinski
tensor introduced in reference [\s]; however, it has been pointed out
that this tensor is identically zero [\wB].  This can been seen very
easily from the Jacobi identity (2.31), since $[I,N] = [I,[I,I]]$.

We can continue to compute commutators
 (or Poisson brackets), but it seems that
this is not a finitely-generated W-algebra [\dn].  However, if the
transformations are restricted to be rigid,
 then the $\hat {N}$ symmetry will not
be generated starting from $\delta_I$ and
 the algebra generated by $\delta_I$ and
$\delta_N$ closes.  This is therefore a finite-dimensional
 rigid symmetry algebra.

In the case of zero Wess-Zumino term $H=0$, this case reduces to N=2
superconformal symmetry.

\section {$SU({n\over 2}) = SU(m)$}

In the case of $SU(m)$ we have, in addition to the almost complex
structure $I$, an m-form $\omega_L$ which is the sum of an (m,0) and a
(0,m) form, $\omega_L = \epsilon + \bar {\epsilon}$.  In a unitary basis
$\epsilon_{a_1...a_m}$ is the usual
 $\epsilon$-tensor in m-dimensions.  We also
have another m-form   $\omega_{\hat {L}} = {1\over i}(\epsilon - \bar
{\epsilon})$. We shall suppose that $I$
 is a complex structure.   There are thus
three transformations to consider, $\delta_I$, $\delta_L$ and $\delta_{\hat
{L}}$, where
$$\omega_L= L_{i_1...i_m}\  dx^{i_1}\wedge\cdots \wedge dx^{i_m}\eqn\bfive$$
$l=(m-1)$ in the notation of
section 2, and $\hat L^i{}_L= I^i{}_j\ L^j{}_L$.
The algebra generated by $\delta_I$ closes as $N = 0$.  In the
commutator of $\delta_I$ with $\delta_L$ one
 finds that the terms involving
 ${I^i} _j\  {{\cal L}^j}_k +{{\cal L}^i} _j\  {I^j}_k$ and
 $[I,L]$ are zero using the fact that $\omega_L= \epsilon +
\bar {\epsilon}$.  .Thus we are left
with
$$[\delta_I,\delta_L]X^i =\
 -\ \big{(}(m-1)\  a_{1-m}\ D_+a_{-1}\  +\  a_{-1}\
D_+a_{1-m}\ \big {)}\  {\hat {L}^i}_L\  D_+X^L  \eqn\bsix$$

In the commutator of $\delta_I$ and $\delta_{\hat {L}}$
the third term vanishes because $\hat {L}$ is the sum of
an $(m,0)$-form and a $(0,m)$-form.  The
first term can be shown to be zero by using the Jacobi identity
(2.32) and the fact that
 the Nijenhuis tensor $[I,L]$ vanishes.  Finally
$$[\delta_I,\delta_{\hat {L}}]X^i= \big{(}(m-1)\  a_{1-m}\ D_+a_{-1}\  +\
a_{-1}\  D_+a_{1-m}\big {)}\  {L^i}_{j_1...j_{m-1}}\
D_+X^{j_1}\cdots D_+X^{j_{m-1}}
				\eqn\beighta$$
closes to a $\delta_L$
transformation     The commutator of two $\delta_L$ transformations yields
$$\eqalign {
[&\delta_L,\delta'_L] X^i
=
a_{1-m}\  a'_{1-m}\  {[L,L]^i}_{j_1...j_{2m-2}}\
D_+X^{j_1}\cdots D_+X^{j_{2m-2}}\ +
 \cr &
+
 \big {(} a'_{1-m}\ D_+a_{1-m}\  -\  a_{1-m}\  D_+a'_{1-m}\big {)}\
{(L\bar {\wedge} L)^i}_{j_1...j_{2m-3}}\  D_+X^{j_1}\cdots D_+X^{j_{2m-3}}\ +
\cr &
+\
i\  (m-1)^2\  a_{1-m}\  a'_{1-m}\  (1-(-1)^{m-1})\
\cr &
({\cal L}^i_l\  {\cal L}^l_k)_{j_1...j_{2m-4}}\   \partial_{\pp}X^k\
D_+X^{j_1}\cdots D_+X^{j_{2m-4}}\cr}    \eqn\beightb$$
For $m$ odd the first and last terms vanish, but the second term
does not vanish for any $m$. In general, therefore, the algebra
generated by $I$ and $L$ is very complicated and leads to an
infinite number of (super)conformal symmetries.  We can get a
finite-dimensional W-algebra by taking the Wess-Zumino $H$  term to vanish.
In this case we recover the W-algebra presented in ref.[\paho].  If in
addition we assume that the parameters of the $\delta_I$ and $\delta_L$ are
rigid and $m$ is an odd number,  $[\delta_L,\delta_L]X^i=0$ and the $\hat {L}$
transformations are not generated as the parameters are restricted to be
constant.  This subset of cases includes six- (real) dimensional Calabi-Yau
spaces.
The Poisson bracket algebra of the currents of the symmetries of sigma models
with target manifold $M$ with $SU(m)$ holonomy and without Wess-Zumino term
closes as W-algebra.   Indeed
$$\eqalign{
\{j_I(a_{-1}), j_I(a'_{-1})\}_{PB}&=\  -\ i\  T(a_{-1}\  a'_{-1})
\cr
\{j_I(a_{-1}), j_L(a_{-l})\}_{PB}&= \ -\ j_{\hat {L}}\big{(}{1 \over l+1}
D_+a_{-1}\  a_{-l}  - D_+(a_{-1}\  a_{-l})\big {)} \cr
\{j_I(a_{-1}), j_{\hat {L}}(a_{-l})\}_{PB}& =\ -\ j_{L}\big {(}{1 \over l+1}\
D_+a_{-1}\  a_{-l}  - D_+(a_{-1}\  a_{-l})\big {)} \cr};$$
For $l$ odd
$$\eqalign {
\{j_L(a_{-l}), j_L(a'_{-l})\}_{PB}&=\ -\  i\  l\ .l!\  T \big {(}a_{-l}\
a'_{-l}\  {j_{I}}^{l-1}\big {)}
\cr
\{j_{\hat {L}}(a_{-l}), j_{\hat {L}}(a'_{-l})\}_{PB}&=\ - \  i\  l\ .l!\  T
\big {(}a_{-l}\ a'_{-l}\  {j_{I}}^{l-1}\big {)}
\cr
\{j_{\hat {L}}(\hat {a}_{-l}), j_L(a_{-l})\}_{PB}&=\ l!\ j_I\ \big {(} \{2
D_+\hat {a}_{-l}\ a_{-l}\  +\  D_+(\hat {a}_{-l}\ a_{-l})\}\  {j_I}^{l-1} \big
{)} \cr }
					\eqn\baeightb$$
and for $l$ even
$$\eqalign{
\{j_L(a_{-l}), j_L(a'_{-l})\}_{PB}&=\  l!\ j_I\big {(}\{2 D_+a_{-l}\
a'_{-l}\  -\  D_+a_{-l}\ a'_{-l})\}\  {j_I}^{l-1}\big {)}
\cr
\{j_{\hat {L}}(a_{-l}), j_{\hat {L}}(a'_{-l})\}_{PB}&=\  l!\  j_I\big {(}\{2
D_+a_{-l}\ a'_{-l}\  -\  D_+a_{-l}\  a'_{-l})\}\  {j_I}^{l-1} \big{)}
\cr
\{j_{\hat {L}}(\hat {a}_{-l}), j_L(a_{-l})\}_{PB}&= - \ i\ l\ l!\ T\big {(}
\hat {a}_{-l}\ a_{-l}\ {j_I}^{l-1}\big {)}  \cr}
				\eqn\baeighta$$

\section {$Sp({n\over 4})=Sp(m)$; $Sp(1)\cdot Sp(m)$}

If the holonomy group can be reduced to $Sp(m)$ there are almost complex
structures $I_r$, r=1,...3, which satisfy the algebra of imaginary unit
quaternions,
$$ I_r\ I_s=-\delta_{rs} + \epsilon_{rst}\  I_t.  \eqn\bsone$$
The corresponding covariantly constant forms are obtained by lowering
an index with the metric, which is hermitian with respect to all three
complex structures.  These structures can be used to define three
additional supersymmetries, in the usual way,
$$\delta_r X^i=\ a^r_{-1}\  {{I_r}^i}_j\  D_+ X^j  \eqn\bstwo$$
The commutator of the algebra closes, except for the terms involving
the Nijenhuis tensors $[I_r,I_s]$.  These generate new symmetries
as in the N=2 case discussed above.

In the case of zero Wess-Zumino term, the algebra of currents of the above
transformations is the N=4 superconformal algebra.

In the case $Sp(1)\cdot Sp(m)$ the three complex structures
are not globally defined on the target space.  The symmetry transformations
(3.11) may be defined only in the case of local supersymmetry [\bw, \wn].
However, there is a covariantly constant four-form $\omega_L$ given by
$$\omega _L = \sum_{r=1}^3 \omega_r \wedge \omega_r   \eqn\bsthree$$
where $\omega_r$ is the two form corresponding to $I_r$.  This can
be used to define a transformation of the type (2.12).  The Poisson algebra
\aasix\ of the current $j_L$ of the corresponding symmetry   is
$$\{ j_L(a_{-3}), j_L(a'_{-3}) \}_{PB} =
\ {  i\over 4}\  j_L ( a_{-3} a'_{-3} T).                      \eqn\batwo$$

\section {$G_2$ and $Spin(7)$}

These two cases are closely related.  We begin with $G_2$.  It
is the subgroup of $SO(7)$ which leaves the antisymmetric
three-index tensor defined by the structure constants of the
imaginary unit octonions invariant.  If $e^a$ is a basis of
orthonormal frames on $M$ the corresponding three-form,
$\varphi$, is
$$\varphi= e^{123} + e^{145} + e^{167} + e^{246} - e^{257} -
e^{356} - e^{347}    \eqn\bnine$$
where
$$e^{abc}=e^a \wedge e^b \wedge e^c.  \eqn\bten$$
We also write
$$\varphi=\omega_L=\ L_{ijk}\  dx^i \wedge dx^j \wedge dx^k.  \eqn\beleven$$
We observe that the covariant constancy of $L_{ijk}$ with
respect to the connection $\Gamma^{(+)}$ implies that the
torsion $H$ must vanish.  The equation of covariant constancy
can be written in the form
$$\nabla_i L_{jkl} - {1\over 2} {H^m} _{i[j}\  L_{kl]m} = 0
\eqn\btwelve$$
where $\nabla$ is the Levi-Civita connection.  Using (3.13) one
observes that
$$\nabla_i L_{jkl} = 0   \eqn\bthirteen$$
and
$${H^m} _{i[j}\  L_{kl]m} = 0    \eqn\bfourteen$$
are valid separately.  Finally one can show that (4.21) implies the
vanishing of $H_{ijk}$.

A second invariant tensor can be defined as the dual of $\omega_L$,
$$^*\omega_L = M_{ijkl}\  dx^i \wedge \cdots \wedge dx^l   \eqn\bfifteen$$
Therefore we have an algebra generated by $\delta_L$ and $\delta_M$.  The
commutator of two $L$-transformations,
 gives an $M$-transformation the parameter
of which vanishes in the rigid case,
$$[\delta_L,\delta'_L]X^i =
-2 (a'_{-2} D_+a_{-2} - a_{-2} D_+a'_{-2})\  M^i_{jkl}\  D_+X^j \cdots D_+X^l
\eqn\bsixteen$$
Thus, if we take the parameters to be constant there is an Abelian symmetry
algebra generated by $L$ alone.

In general, the Poisson bracket algebra of these symmetries closes as a W
algebra. Indeed,
$$\eqalign {
\{ j_L(a_{-2}), j_L(a'_{-2}) \}_{PB} &= - 2\  j_M\big {(} 2\ D_+a_{-2}\
a'_{-2} - D_+(a_{-2}\  a'_{-2}) \big {)}
\cr
\{ j_M(a_{-3}), j_L(a_{-2}) \}_{PB} &=\  27\  i\  j_L\big {(} a_{-3}\  a_{-2}\
T\big {)}
 \cr
\{ j_M(a_{-3}), j_L(a'_{-3}) \}_{PB} &=\  {9 i\over 4}\   j_M\big {(} a_{-3}\
a'_{-3}\  T\ \big {)}\  -\
 9\ j_L\big {(} a_{-3}\  a'_{-3}\  D_+j_L\ \big {)}
	\cr}                 \eqn\basixteen$$

 Finally, we turn to $Spin(7)$. The
 target manifold in this case has dimension
$8$ and the invariant tensor is a self-dual $4$-form $\Phi$ which can be
constructed from  $\varphi$.  Let $e^0$, $e^a$, $a=1,\cdots,7$, be an
orthonormal basis, then
$$\eqalign {\Phi &= e^0 \wedge \varphi + ^*\varphi;
\cr
\Phi=\omega_L &=
L_{ijkl}\ dx^i \ \wedge \cdots \wedge dx^l
				\cr}  \eqn\bseventeen$$
It is straightforward to verify
 that the torsion $H$ vanishes in the $Spin(7)$
case as it does in the $G_2$ case.  The Poisson bracket algebra of two
tranformations
 generated by $\omega_L$ closes as W algebra; it is
$$ \{ j_L(a_{-3}), j_L(a'_{-3}) \}_{PB} =\  {9 i\over 4}\  j_L\big {(}a_{-3}\
a'_{-3}\  T\ \big {)}  		\eqn\beighteen$$

 \chapter {Concluding Remarks}

In this paper we have seen that two-dimensional supersymmetric sigma models
on Riemannian target spaces with
 special geometries have associated symmetries
and that, classically, the algebraic structure of these symmetries is of
W-type, i.e. higher spin extensions of the superconformal algebra. It would
clearly be of interest to analyse these symmetries at the quantum level, but,
as we remarked in the introduction, this is non-trivial in view of the
non-linearities involved. If one makes the assumption that symmetries of this
type are preserved quantum mechanically,
 then they would seem, in certain cases,
to imply strong constraints on the renormalisation of the models concerned.
For example, $N=1$ sigma models on Calabi-Yau target spaces have additional
symmetries of this type as we have seen, and these, if preserved, would imply,
in conjunction with the Calabi-Yau theorem,
 the perturbative finiteness of such
models. Since this would contradict explicit
 calculations [\gris] (except for $n$=
4), the conclusion seems to be that
 these symmetries are in general anomalous
quantum mechanically. We have carried out a prelimanary calculation for the
case $n=6$ which lends support to this conjecture, but a complete analysis
remains to be done.

\refout

\end